\documentclass{article}
\usepackage[cp866]{inputenc}
\begin{document}

\centerline {{\Large\bf The connection between field-theory and }}
\centerline {{\Large\bf the equations for material sistems}}
\centerline {\bf L.~I. Petrova}     
\centerline{{\it Moscow State University, Russia, e-mail: ptr@cs.msu.su}}

\renewcommand{\abstractname}{Abstract}
\begin{abstract}
The existing field theories are based on the properties of closed
exterior forms, which correspond to conservation laws for physical 
fields. 

In the present paper it is shown that closed exterior forms
corresponding to field theories are obtained from the equations
modelling conservation (balance) laws for material sistems (material 
media).

The process of obtaining closed exterior forms demonstrates the 
connection between field-theory equations and the equations for material 
sistems and points to the fact that the foundations of field theories 
must be  conditioned by the properties of equations 
conservation laws for material sistems. 

\end{abstract}

\subsection*{1. Peculiarities of differential equations for material 
sistems }

Equations for material sistems are
equations  that describe the conservation laws for energy, linear
momentum, angular momentum and mass. Such conservation laws can be
named as balance ones since they establish the balance  between
the variation of a physical quantity and corresponding external
action.

{\footnotesize [The material system  - material (continuous) medium 
- is a variety (infinite) of elements  that have internal
structure and interact among themselves. Thermodynamical,
gasodynamical and cosmologic system, systems of elementary
particles and others are examples of material system. (Physical
vacuum can be considered as an analog of such material system.)
Electrons, protons, neutrons, atoms, fluid particles and so on are
examples of elements of material system.]}

The equations of balance conservation laws are differential 
(or integral) equations that describe a variation of functions
corresponding to physical quantities [1-3]. (The Navier-Stokes equations are
an example [3].)

It appears that, even without a knowledge of the concrete form of these
equations, one can see specific features of these equations and their
solutions using skew-symmetric differential forms [4-6]. 

To do so it is necessary to study the conjugacy (consistency) of
these equations.

The functions for equations of material sistems sought are usually
functions which relate to such physical quantities like a particle
velocity (of elements), temperature or energy, pressure and
density. Since these functions relate to one material system, it
has to exist a connection between them. This connection is
described by the state-function. Below it will be shown that the
analysis of integrability and consistency of equations of balance
conservation laws for material media reduces to a study the
nonidentical relation for the state-function.

Let us analyze the equations that describe the balance conservation
laws for energy and linear momentum.

We introduce two frames of reference: the first is an inertial one
(this frame of reference is not connected with the material system), and
the second is an accompanying
one (this system is connected with the manifold built by
the trajectories of the material system elements).

The energy equation
in the inertial frame of reference can be reduced to the form:
$$
\frac{D\psi}{Dt}=A_1
$$
where $D/Dt$ is the total derivative with respect to time, $\psi $ is
the functional of the state that specifies the material system, $A_1$ is
the quantity that
depends on specific features of the system and on external energy
actions onto the system.
{\footnotesize \{The action functional, entropy, wave function
can be regarded as examples of the functional $\psi $. Thus, the
equation for energy presented in terms of the action functional $S$ has
a similar form:
$DS/Dt\,=\,L$, where $\psi \,=\,S$, $A_1\,=\,L$ is the Lagrange function.
In mechanics of continuous media the equation for
energy of an ideal gas can be presented in the form [3]: $Ds/Dt\,=\,0$, 
where $s$ is entropy.\}}

In the accompanying frame of reference the total derivative with respect to
time is transformed into the derivative along the trajectory. Equation
of energy is now written in the form
$$
{{\partial \psi }\over {\partial \xi ^1}}\,=\,A_1 \eqno(1)
$$
Here  $\xi^1$ is the coordinate along the trajectory.

In a similar manner, in the accompanying reference system the
equation for linear momentum appears to be reduced to the equation of
the form
$$
{{\partial \psi}\over {\partial \xi^{\nu }}}\,=\,A_{\nu },\quad \nu \,=\,2,\,...\eqno(2)
$$
where $\xi ^{\nu }$ are the coordinates in the direction normal to the
trajectory, $A_{\nu }$ are the quantities that depend on the specific
features of material system and on external force actions.

Eqs. (1) and (2) can be convoluted into the relation
$$
d\psi\,=\,A_{\mu }\,d\xi ^{\mu },\quad (\mu\,=\,1,\,\nu )\eqno(3)
$$
where $d\psi $ is the differential
expression $d\psi\,=\,(\partial \psi /\partial \xi ^{\mu })d\xi ^{\mu }$.

Relation (3) can be written as
$$
d\psi \,=\,\omega \eqno(4)
$$
here $\omega \,=\,A_{\mu }\,d\xi ^{\mu }$ is the skew-symmetrical differential
form of the first degree (the summation over repeated indices is implied).

Relation (4) has been obtained from the equation of the balance conservation
laws for energy and linear momentum. In this relation the form $\omega $
is that of the first degree. If the equations of the balance conservation
laws for angular momentum be added to the equations for energy and linear
momentum, this form  will be a form of the
second degree. And in combination with the equation of the balance
conservation law for mass this form will be a form of degree 3.
In general case the evolutionary relation can be written as
$$
d\psi \,=\,\omega^p \eqno(5)
$$
where the form degree  $p$ takes the values $p\,=\,0,1,2,3$.
{\footnotesize (The relation for $p\,=\,0$ is an analog to
that in the differential forms of zero degree, and it was obtained from the
interaction of energy and time.)}

Since the balance conservation laws are evolutionary ones, the relations
obtained are also evolutionary relations, and the skew-symmetric forms
$\omega $ and $\omega^p $ are evolutionary ones.

Relations obtained from the equation of the balance conservation
laws turn out to be nonidentical.
In the left-hand side of evolutionary relation (4) there is a
differential that is a closed form. This form is an invariant object. 
The right-hand side of relation (4) involves the differential form
$\omega$, that is not an invariant object because in real processes, as 
it is shown below, this form proves to be unclosed.

For the form to be closed the differential of the form or its commutator
must be equal to zero.
Let us consider the commutator of the form 
$\omega \,=\,A_{\mu }d\xi ^{\mu }$.
The components of the commutator of such a form can be written as follows:
$$
K_{\alpha \beta }\,=\,\left ({{\partial A_{\beta }}\over {\partial \xi ^{\alpha }}}\,-\,
{{\partial A_{\alpha }}\over {\partial \xi ^{\beta }}}\right )\eqno(6)
$$
(here the term  connected with the manifold metric form
has not yet been taken into account).

The coefficients $A_{\mu }$ of the form $\omega $ have been obtained either
from the equation of the balance conservation law for energy or from that for
linear momentum. This means that in the first case the coefficients depend
on the energetic action and in the second case they depend on the force action.
In actual processes energetic and force actions have different nature and appear
to be inconsistent. The commutator of the form $\omega $ constructed from
the derivatives of such coefficients is nonzero.
This means that the differential of the form $\omega $
is nonzero as well. Thus, the form $\omega$ proves to be unclosed and is 
not an invariant quantity.

This means that the relation (4) involves not an invariant term. 
Such a relation cannot be an identical one.
Hence, without the knowledge of particular expression for the form
$\omega$, one can argue that for actual processes the relation obtained 
from the equations corresponding to the balance conservation laws proves 
to be nonidentical.
In similar manner it can be shown that  general relation (5) is also
nonidentical.

{\footnotesize \{The peculiarities of the evolutionary relation are 
connected with
the differential form that enters into this relation. This is a
skew-symmetric form with the basis, in contrast to the basis of
exterior form, is a deforming (nondifferentiable) manifold. (About
the properties of such skew-symmetric form one can read, for
example, in paper [6]). The peculiarity of skew-symmetric forms
defined on such manifold is the fact that their differential
depends on the basis. The commutator of such form includes the
term that is connected with a differentiating the basis. This can
be demonstrated by an example of the first-degree skew-symmetric
form.

Let us consider the first-degree form
$\omega=a_\alpha dx^\alpha$. The differential of this form can
be written as $d\omega=K_{\alpha\beta}dx^\alpha dx^\beta$, where
$K_{\alpha\beta}=a_{\beta;\alpha}-a_{\alpha;\beta}$ are
the components of the commutator of the form $\omega$, and
$a_{\beta;\alpha}$, $a_{\alpha;\beta}$ are the covariant
derivatives. If we express the covariant derivatives in terms of
the connectedness (if it is possible), then they can be written
as $a_{\beta;\alpha}=\partial a_\beta/\partial
x^\alpha+\Gamma^\sigma_{\beta\alpha}a_\sigma$, where the first
term results from differentiating the form coefficients, and the
second term results from differentiating the basis. If we substitute
the expressions for covariant derivatives into the formula for
the commutator components, we obtain the following expression
for the commutator components of the form $\omega$:
$$
K_{\alpha\beta}=\left(\frac{\partial a_\beta}{\partial
x^\alpha}-\frac{\partial a_\alpha}{\partial
x^\beta}\right)+(\Gamma^\sigma_{\beta\alpha}-
\Gamma^\sigma_{\alpha\beta})a_\sigma
$$
Here the expressions
$(\Gamma^\sigma_{\beta\alpha}-\Gamma^\sigma_{\alpha\beta})$
entered into the second term are just components of commutator of
the first-degree metric form that specifies the manifold
deformation and hence equals nonzero. (In commutator of the 
exterior form, which is defined on differentiable manifold 
the second term is not present.) [It is well-known that 
the metric form commutators of the first-, second- and third 
degrees specifies, respectively,  torsion, rotation and
curvature.] 

The skew-symmetric form in the evolutionary relation is defined in
the manifold made up by trajectories of the material system
elements. Such a manifold is a
deforming manifold. The commutator of skew-symmetric form  defined on
such manifold includes the metric form commutator being nonzero. 
(The commutator of unclosed metric form, which is nonzero, enters into 
commutator (6) of the evolutionary form 
$\omega \,=\,A_{\mu }d\xi ^{\mu }$.)
Such commutator of differential form cannot vanish. 
And this means that evolutionary skew-symmetric form 
that enters into evolutionary relation cannot be closed.
The evolutionary relation cannot be an identical one.
(Nonclosure of evolutionary form and an availability of additional term
in this form commutator governs the properties and peculiarities of nonidentical
evolutionary relation and its physical importance.)\}} 

Nonidentity of the evolutionary relation means that initial equations 
of balance conservation laws are not conjugated, and hence they are not 
integrable. The solutions of these equations can be functional or 
generalized ones. In this case generalized solutions are obtained only 
under degenerated transformations.

The evolutionary relation obtained from equations of balance
conservation laws for material systems (continuous media) carries not 
only mathematical but also large physical loading [6,7]. This is due to 
the fact that the 
evolutionary relation possesses the duality. On the one hand, this 
relation corresponds to material system, and on other, as it will be 
shown below, describes the mechanism of generating physical structures. 
This discloses the properties and peculiarities of the
field-theory equations and their connection with the equations of
balance conservation laws for material systems.  

\bigskip

{\bf Physical significance of nonidentical evolutionary relation.}

The evolutionary relation describes the evolutionary process in material 
system since this relation includes the state differential $d\psi $, 
which specifies the material system state.
However, since this relation turns out to be not identical, from this
relation one cannot get the differential $d\psi $. The absence of
differential means that the system state is nonequilibrium.

The evolutionary relation possesses one more peculiarity,
namely, this relation is a selfvarying relation. (The evolutionary form
entering into this relation is defined on the deforming manifold
made up by trajectories of the material system elements. This means
that the evolutionary form basis varies. In turn,
this leads to variation of the evolutionary form, and the process
of intervariation of the evolutionary form and the basis is
repeated.)

Selfvariation of the nonidentical evolutionary relation points to the
fact that the nonequilibrium state of material system turns out
to be selfvarying. {\footnotesize (It is evident that this selfvariation proceeds under the action of
internal force whose quantity is described by the commutator of the
unclosed evolutionary form $\omega^p $.)}
State of material system changes but remains
nonequilibrium during this process.

Since the evolutionary form is unclosed, the evolutionary relation cannot be
identical. This means that the nonequilibrium state of material
system holds. But in this case it is possible a transition of material
system to a locally equilibrium state.

This follows from one more property of nonidentical
evolutionary relation. Under selfvariation of the evolutionary relation
it can be realized the conditions of degenerate transformation.
And under degenerate transformation from the nonidentical relation it
is obtained the identical relation.

From identical relation one can define the state differential pointing
to the equilibrium state of the system. However, such system state is
realized only locally due to the fact that the state differential
obtained is an interior one defined only on pseudostructure, that is
specified by the conditions of degenerate transformation. And yet
the total state of material system remains to be
nonequilibrium because the evolutionary relation, which describes
the material system state, remains nonidentical one.

The conditions of degenerate transformation are connected with symmetries 
caused by degrees of freedom of material system. These are symmetries of
the metric forms commutators of the manifold.  {\footnotesize \{To the
degenerate transformation it must correspond a vanishing of some
functional expressions, such as Jacobians, determinants, the Poisson
brackets, residues and others. Vanishing of these
functional expressions is the closure condition for dual form.
And it should be emphasize once more that the degenerate
transformation is realized as a transition from the accompanying
noninertial frame of reference to the locally inertial system. The
evolutionary form and nonidentical evolutionary relation are defined in
the noninertial frame of reference (deforming manifold). But the closed
exterior form obtained and the identical relation are obtained
with respect to the locally-inertial frame of reference
(pseudostructure)\}}.

Realization of the conditions of degenerate transformation is a vanishing
of the commutator of manifold metric form, that is, a vanishing of
the dual form commutator. And this leads to realization of pseudostructure
and formatting the closed inexact form, whose closure conditions have the form
$$d_\pi \omega^p=0,  d_\pi{}^*\omega^p=0 \eqno(7)$$
On the pseudostructure $\pi$ from evolutionary relation (5) it is 
obtained the relation
$$
d_\pi\psi=\omega_\pi^p\eqno(8)
$$
which proves to be an identical relation  since the closed inexact form
is a differential (interior on pseudostructure).

The realization of the conditions of degenerate transformation and
obtaining identical relation from nonidentical one has both
mathematical and physical meaning. Firstly, this points to the
fact that the solution of equations of balance conservation laws
proves to be a generalized one. And secondly, from this relation
one obtains the differential $d_\pi\psi $ and this points to the
availability of the state-function (potential) and that the state
of material system is in local equilibrium.

\bigskip
Relation (8) holds the duality. The left-hand side of relation
(8) includes the differential, which specifies material system
and whose availability points to the locally-equilibrium state of
material system. And the right-hand side includes a closed inexact
form, which is a characteristics of physical fields. The closure
conditions (7) for exterior inexact form correspond to the
conservation law, i.e. to a conservative on pseudostructure
quantity, and describe a differential-geometrical structure. These
are such structures (pseudostructures with conservative
quantities) that are physical structures formatting physical
fields[6,7].

The transition from nonidentical relation (5) obtained from equations of 
the balance conservation laws to identical
relation (8) means the following. Firstly, an emergency of the
closed (on pseudostructure) inexact exterior form (right-hand side
of relation (8)) points to an origination of the physical structure.
And, secondly, an existence of the state differential (left-hand side
of relation (8)) points to a transition of the material system from nonequilibrium state
to the locally-equilibrium state.

Thus one can see that the transition of material system from
nonequilibrium state to locally-equilibrium state is accompanied
by originating differential-geometrical structures, which are
physical structures.  Massless particles, charges,
structures made up by eikonal surfaces and wave fronts, and so on are
examples of physical structures.

The duality of identical relation also explains the duality of 
nonidentical evolutionary relation. On the one hand, evoltionary 
relation describes the evolutionary process in material systems, 
and on the other describes the process of generating physical fields.

The emergency of physical structures in the evolutionary process
reveals in material system as an emergency of certain observable
formations, which develop spontaneously. Such formations and their
manifestations are fluctuations, turbulent pulsations, waves, vortices,
and others. It appears that structures of physical fields and the 
formations of material systems observed are a manifestation of the same 
phenomena. The light is an example of such a duality. The light 
manifests itself in the form of a massless particle (photon) and of 
a wave.

This duality also explains a distinction in studying the same phenomena
in material systems and physical fields. In the physics of continuous 
media (material systems) the interest is expressed
in generalized solutions of equations of the balance conservation laws.
These are solutions that describe the formations in material media 
observed. The investigation of relevant physical structures is carried 
out using the field-theory equations.

\bigskip
The unique properties of nonidentical evolutionary relation, which
describes the connection between physical fields and material
systems, discloses the connection of evolutionary relation with
the field-theory equations. In fact, all equations of existing
field theories are the analog to such relation or its differential
or tensor representation.

\subsection*{2. Specific features of field-theory equations}

The field-theory equations are equations that describe physical
fields. Since physical fields are formatted by physical structures,
which are described by closed exterior {\it inexact } forms and
by closed dual forms (metric forms of manifold), is obvious that
the field-theory equations or solutions to these equations have to
be connected with closed exterior forms. Nonidentical relations
for functionals like wave-function, action functional, entropy,
and others, which are obtained from the equations for material
media (and from which identical relations with closed forms
describing physical fields are obtained), just disclose the
specific features  of the field-theory equations.

The equations of mechanics, as well as the equations of continuous
media physics, are partial differential equations for desired
functions like a velocity of particles (elements), temperature,
pressure and density, which correspond to physical quantities of
material systems (continuous media). Such functions describe the
character of varying physical quantities of material system. The
functionals (and state-functions) like wave-function, action
functional, entropy and others, which specify the state of
material systems, and corresponding relations are used in
mechanics and continuous media physics only for analysis of
integrability of these equations. And in field theories such
relations play a role of equations. Here it reveals the duality of
these relations. In mechanics and continuous media physics these
equations describe the state of material systems, whereas in
field-theory they describe physical structures from which physical
fields are formatted.

\bigskip
It can be shown that all equations of existing field theories are
in essence relations that connect skew-symmetric forms or their analogs
(differential or tensor ones). And yet
the nonidentical relations are treated as equations from which it can be
found identical relation with include closed forms
describing physical structures desired.

Field equations (the equations of the Hamilton formalism) reduce
to identical relation with exterior form of first degree, namely, 
to the Poincare invariant $$ds\,=-\,H\,dt\,+\,p_j\,dq_j\eqno(9)$$

{\footnotesize \{As is known, the field equation has the form 
$${{\partial s}\over {\partial t}}+H \left(t,\,q_j,\,{{\partial s}\over {\partial q_j}}
\right )\,=\,0,\quad
{{\partial s}\over {\partial q_j}}\,=\,p_j \eqno(10)$$
here $s$ is a field function for the action functional $S\,=\,\int\,L\,dt$.
Here $L$ is the Lagrangian function, $H(t,\,q_j,\,p_j)\,=\,p_j\,\dot q_j-L$
is the Hamilton function $p_j\,=\,\partial L/\partial \dot q_j$. These functions
satisfy the relations:
$${{dg_j}\over {dt}}\,=\,{{\partial H}\over {\partial p_j}}, \quad
{{dp_j}\over {dt}}\,=\,-{{\partial H}\over {\partial g_j}}\eqno(11)$$
Relations (11), which present a set of the Hamilton equations, are the closure
conditions for exterior and dual forms [6].\}}

The Schr\H{o}dinger equation in quantum mechanics is an analog to
field equation, where the conjugated coordinates are replaced by
operators. The Heisenberg equation corresponds to the closure
condition of dual form of zero degree. Dirac's {\it bra-} and {\it cket}- vectors made
up a closed exterior form of zero degree.
It is evident that the relations with skew-symmetric differential forms
of zero degree correspond to quantum mechanics.
The properties of skew-symmetric differential forms
of the second degree lie at the basis of the electromagnetic field
equations. The Maxwell equations may be written as $d\theta^2=0$, $d^*\theta^2=0$,
where $\theta^2= \frac{1}{2}F_{\mu\nu}dx^\mu dx^\nu$ (here
$F_{\mu\nu}$ is the strength tensor). The Einstein equation is a
relation in differential forms. This equation relates the
differential of dual form of first degree (Einstein's tensor) and
a closed form of second degree --the energy-momentum tensor. (It
can be noted that, Einstein's equation is obtained from
differential forms of third degree).

The connection the field theory equations with skew-symmetric
forms of appropriate degrees shows that there exists a commonness
between field theories describing physical fields of different
types. This can serve as an approach to constructing the unified
field theory. This connection shows that it is possible to
introduce a classification of physical fields according to the
degree of skew-symmetric differential forms. From relations  (5)
and (8) one can see that relevant degree of skew-symmetric
differential forms, which can serve as a parameter of unified
field theory, is connected with the degree $p$ of evolutionary
form in relation (5). It should be noted that the degree $p$ is
connected with the number of interacting balance conservation
laws. {\footnotesize \{The degree of closed forms also reflects a
type of interaction [7]. Zero degree is assigned to a strong
interaction, the first one does to a weak interaction, the second
one does to electromagnetic interactions, and the third degree is
assigned to gravitational field.\}}

The connection of field-theory equations, which describe physical fields,
with the equations for material media discloses the foundations of the
general field theory. As an equation of general field theory it can 
serve the evolutionary relation (5), which is obtained from equations 
the balance conservation laws for material system and has a double 
meaning. On the one hand, that, being a relation, specifies the type of 
solutions to equations of balance conservation
laws and describes the state of material system (since it includes
the state differential), and, from other hand, that can play a role
of equations for description of physical fields (for finding the closed
inexact forms, which describe the physical structures from which physical
fields are made up). It is just a double meaning that discloses the
connection of physical fields with material media (which is based on the
conservation laws) and allows to understand on what the general field 
theory has to be based.

\bigskip

In conclusion it should be emphasized that the study of equations
of mathematical physics appears to be possible due to unique
properties of skew-symmetric differential forms. In this case,
beside the exterior skew-symmetric differential forms, which 
are defined on differentiable manifolds, the skew-symmetric differential 
forms, which, unlike to the exterior forms, are defined on deforming 
(nondifferentiable) manifolds [6], were used.

1. Tolman R.~C., Relativity, Thermodynamics, and Cosmology. Clarendon Press, 
Oxford,  UK, 1969.

2. Fock V.~A., Theory of space, time, and gravitation. -Moscow, 
Tech.~Theor.~Lit., 1955 (in Russian).
      
3. Clark J.~F., Machesney ~M., The Dynamics of Real Gases. Butterworths, 
London, 1964. 
 
4. Cartan E., Les Systemes Differentials Exterieus ef Leurs Application 
Geometriques. -Paris, Hermann, 1945.  

5. Schutz B.~F., Geometrical Methods of Mathematical Physics. Cambrige 
University Press, Cambrige, 1982.

6. Petrova L.~I. Skew-symmetric differential forms: Conservation laws. 
Foundations of field theories. -Moscow, URSS, 2006, 158 p. (in Russian). 

7. Petrova L.~I. The quantum character of physical fields. Foundations 
of field theories, Electronic Journal of Theoretical Physics, v.3, 10 
(2006), 89-107p.

\end{document}